# Magnetic Properties of the Heusler Ru2MnX (X = Nb, Ta or V) Compounds: Monte Carlo Simulations

N. Saber, Z. Fadil, A. Mhirech, B. Kabouchi, L. Bahmad* and W. Ousi Benomar

Laboratoire de Matière Condensée et Sciences Interdisciplinaires (LaMCScI), Faculty of Sciences. P.O. Box 1014, Mohammed V University in Rabat, Morocco

**Abstract:**

In this paper, we have focused on a comparison of the different magnetic properties of the three nano-Heusler Ru2MnX (X = Nb, Ta or V) compounds using the Blume-Capel Ising model. The Heusler structures are composed by different mixed spins. In fact, the Ru and Mn atoms are modeled by spin-5/2 and spin-1/2, respectively. While, the X atoms (X = Nb, Ta and V) are represented by the spin-7/2, spin-3/2 and spin-5/2, respectively.

This study is carried out by using the Monte Carlo simulations under the Metropolis algorithm. The magnetic behaviors of the three nano-Heusler compounds have been studied and discussed. It is found that Ferrimagnetic to superparamagnetic transitions were observed corresponding to different blocking temperatures. Besides, the effect of the crystal field, the exchange coupling interactions and the external magnetic field have been inspected on the magnetization of each nano-Heusler compound Ru2MnX (X = Nb, Ta or V).

---

*) Corresponding author: bahmad@fsr.ac.ma (L. B.)

## 1. Introduction:

Heusler alloys [1,2] have become one of the important groups of materials for rapidly growing technology due to their application [3-10]. These materials are often used in spintronics, shape memory alloys, magnetocaloric materials and many others [11]. Frequently, Heusler alloys can be classified into four main groups as alloys of full Heusler, half Heusler, reverse Heusler and quaternary Heusler [11]. These Heusler alloy groups have received great interest especially through theoretical work [12-17]. Moreover, Monte Carlo simulations (MCS) have made it possible to study the magnetic or dielectric properties of different structures by determining and varying the transition temperature in Ising ladder-like graphene nanoribbon [18], bilayer graphene-like structure [19], borophene layers structure with RKKY interactions [20], bi-layer graphyne-like structure [21], monolayer coronene-like nano-structure [22], monolayer nano-graphyne structure [23], diamond-like decorated square [24] and in decorated triangular lattice [25]. The MCS are also used to investigate the compensation temperature, in spin-1⁄2 Ising trilayer [26], trilayer graphyne-like [27], monolayer naphthalene-like nanoisland [28], borophene core-shell [29], ovalene [30] and in nano-trilayer graphene [31]. Indeed, MCS leads to interesting results and allows a better understanding of the nanostructures by varying different physical parameters. Furthermore, the MCS has been used to investigate the magnetic properties of the different Heusler compounds in Co-Fe-Mn-Si [32], Ni-Mn-Ga [33], Ni-Mn-Ga-Cr [34] and magnetocaloric effect in $Ni_{50}Mn_{34}In_{16}$ [35]. In some our earlier works, we have studied and discussed the different properties of the different Heusler structures [36-39].

In the present work, we study the magnetic properties of Heusler compounds $Ru_2MnX$ (X = Nb, Ta or V). Indeed, we aim is to compare the behavior magnetic properties in these three Heusler compounds. The investigated system is explored using the well-known Monte Carlo simulations under the Metropolis algorithm. As far as we know, there are no theoretical investigations in the literature that have studied the magnetic properties of Heusler compounds $Ru_2MnX$ (X = Nb, Ta or V). This paper is organized as follows: the model and method used are illustrated in section 2, the numerical results and discussion are reported in section 3. Finally, a conclusion is given in section 4.

## 2. Model and method:

In this work, we study the magnetic properties of nano-Heusler structures $Ru_2MnX$ composed with three types of atoms: Ruthenium, Manganese and X. In our case, we study the nano-Heusler structure by replacing X atom with three types of atoms: titanium, vanadium and niobium (X=Ta,V or Nb), respectively.

In this work, the total number of each atom is as follows: $N_{Ru}=123$, $N_{Mn}=32$ and $N_{X=Ta, V, Nb} =32$ (see Fig. 1). Besides, such nano-Heusler structures are studied using the Blume Capel model with free boundary conditions. Additionally, our data were generated with $10^6$ Monte Carlo steps per spin, we neglect the $10^5$ first Monte Carlo steps of the simulations to balancing the system.

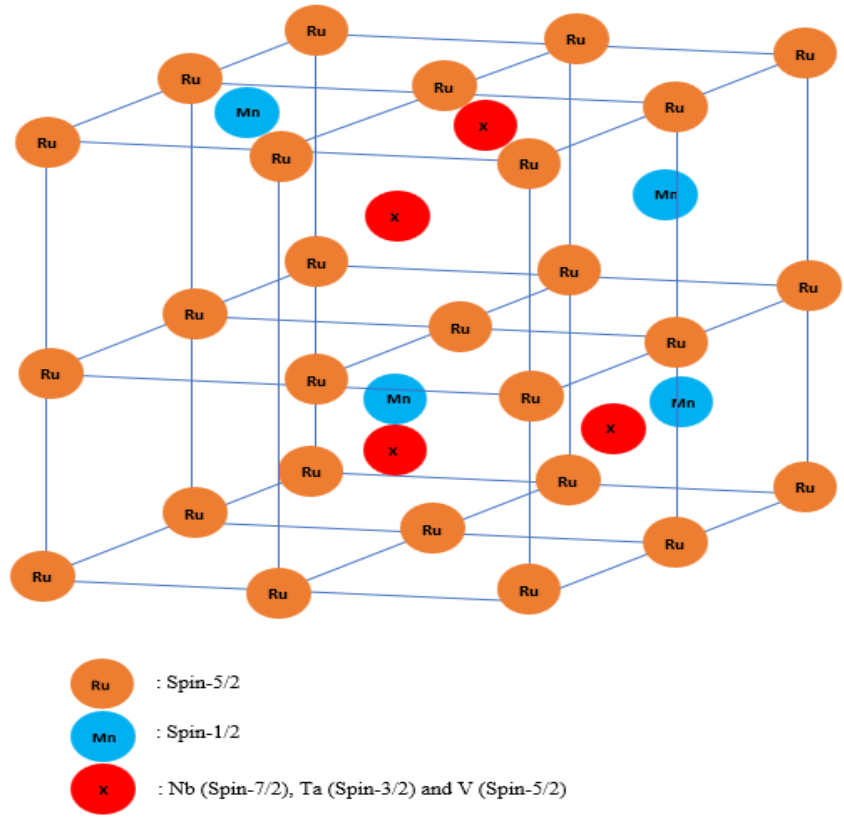

**Fig. 1**: A schematic representation of Heusler structure $Ru_2MnX$ (X=Ta,V or Nb), Ru with spin-5/2 (orange balls), Mn with spin-1/2 (blue balls), Ta with spin-7/2 , V with spin-3/2 and Nb with spin-5/2 (red balls).

The Hamiltonian of Heusler structure is defined as:

$$\mathcal{H} = -J_{Ru\,Mn} \sum_{<i,j>} Ru_i Mn_j - J_{Ru\,X} \sum_{<i,k>} Ru_i X_k - H \sum_i (Ru + Mn + X)$$
$$-D_{Ru} \sum_i Ru_i^2 - D_X \sum_k X_k^2 \qquad (1)$$

where $J_{Ru\,Mn}$, and $J_{Ru\,X}$ represents the exchange coupling interactions between two first nearest neighbor atoms with spins $Ru - Mn$, $Ru - X$ (X=Ta, V or Nb), respectively. H is the external magnetic field. The crystal fields $D_{Ru}$ and $D_X$ are acting on the spin atoms of Ru and X, respectively. In all this work, we will take the identical crystal fields acting on the spins Ru and X: $D = D_{Ru} = D_X$.

The internal energy per site of the studied Heusler structure is:

$$E = \frac{1}{N_T} \langle \mathcal{H} \rangle \qquad (2)$$

With: $N_T = N_{Ru} + N_{Mn} + N_{X=Ta,\,V,\,Nb} = 123 + 32 + 32 = 187$ atoms

The corresponding partial and total magnetizations are:

$$M_{Ru} = \frac{1}{N_{Ru}} \sum_i Ru_i \qquad (3)$$

$$M_{Mn} = \frac{1}{N_{Mn}} \sum_j Mn_j \qquad (4)$$

$$M_X = \frac{1}{N_X} \sum_k X_k \qquad (5)$$

With $N_X = N_{Ta} = N_V = N_{Nb}$

$$M_{tot} = \frac{N_{Ru} M_{Ru} + N_{Mn} M_{Mn} + N_X M_X}{N_{Ru} + N_{Mn} + N_X} \qquad (6)$$

## 3. Numerical results and discussion

In this section, we study the magnetic properties of three nano-Heusler structures Ru$_2$MnX (X=Ta, V or Nb) using Monte Carlo simulations under the Metropolis algorithm.

In Fig. 2a, 2b and 2c, we illustrate the magnetizations of different nano-Heusler structures as a function of the temperature: Ru$_2$MnTa, Ru$_2$MnV and Ru$_2$MnNb, respectively. These figures are plotted in the absence of the crystal and external magnetic fields (D=0 and H=0) and for fixed exchange coupling interactions: J$_{Ru-Mn}$ = 1 and J$_{Ru-X}$ = -1.

The magnetizations at very low temperature (t≈0) are M$_{Ru}$ = -5/2, M$_{Mn}$ = -1/2, M$_{Ta}$ = +7/2, M$_V$ =+3/2 and M$_{Nb}$ = +5/2 leading to M$_{tot}$= $\frac{123\times(-5/2)+32\times(-1/2)+32\times(+7/2)}{(123+32+32)}$ = - 0.5 for Ru$_2$MnTa, M$_{tot}$=$\frac{123\times(-5/2)+32\times(-1/2)+32\times(+3/2)}{(123+32+32)}$ = - 1.3 for Ru$_2$MnV and M$_{tot}$= $\frac{123\times(-5/2)+32\times(-1/2)+32\times(+5/2)}{(123+32+32)}$ = - 0.5 for Ru$_2$MnNb.

Moreover, in Figs. 2a, 2b and 2c, the total magnetizations of Ru2MnTa, Ru2MnV and Ru2MnNb systems are in the ferrimagnetic phase where the temperature of these systems is lower than the blocking temperature (T<T$_B$), then the magnetizations increase when increasing the temperature values to reach the superparamagnetic phase (T> T$_B$). Additionally, to confirm this result and find the exact value of the blocking temperature which determines the transition between the ferrimagnetic and superparamagnetic phase, we have plotted in Fig. 2d, 2e and 2f, the total susceptibilities of the three studied systems. Indeed, the blocking temperatures of the studied systems are located at the peaks of the total susceptibilities T$_B$ ≈ 16.25, 12.5 and 15 for R$_{u2}$MnTa, R$_{u2}$MnV and R$_{u2}$MnNb, respectively.

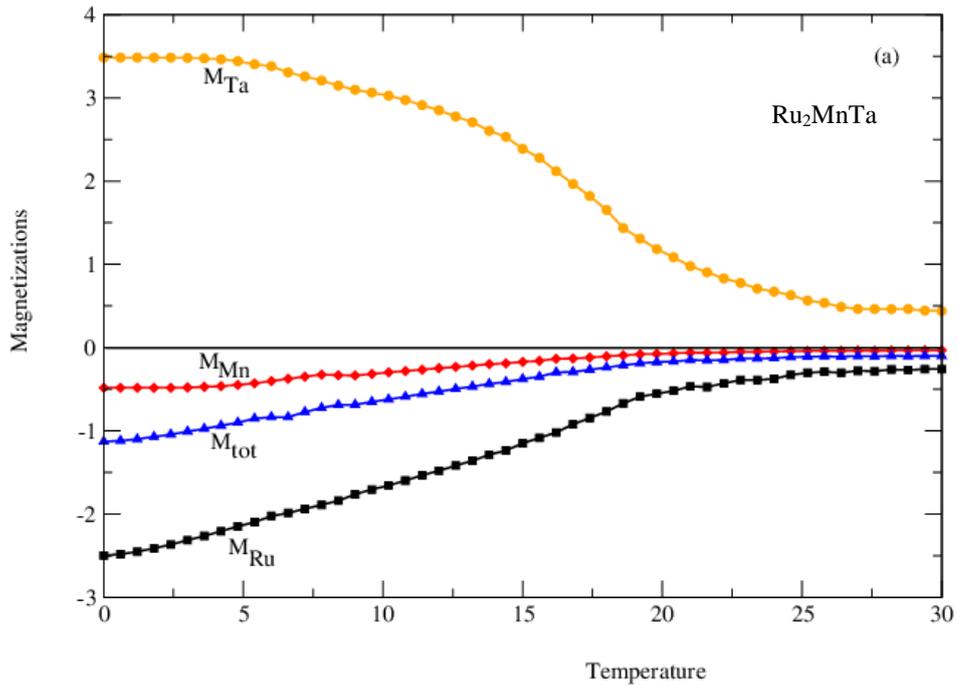

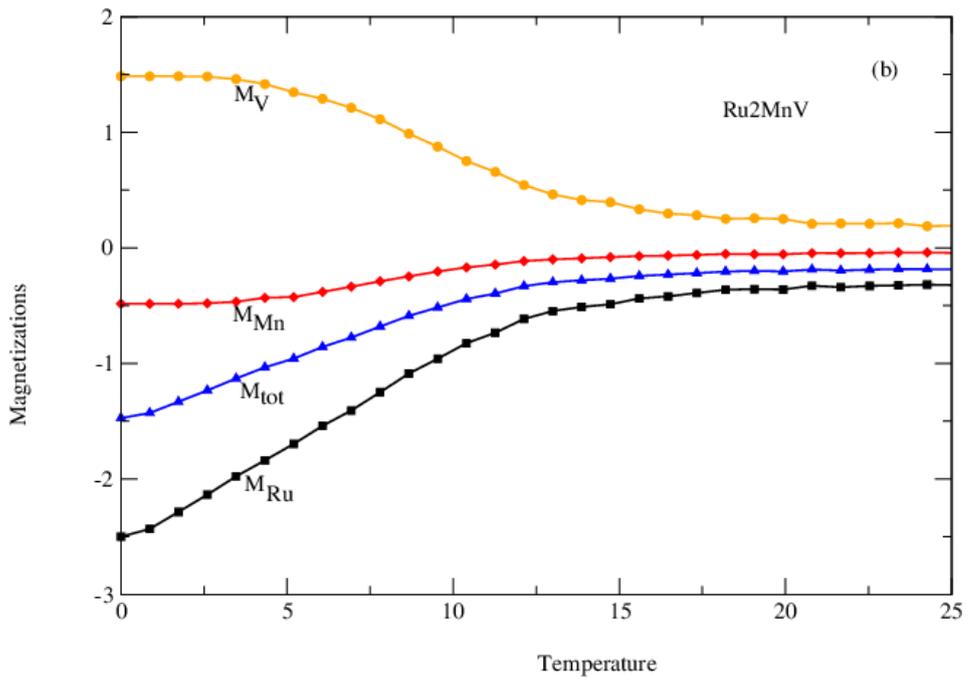

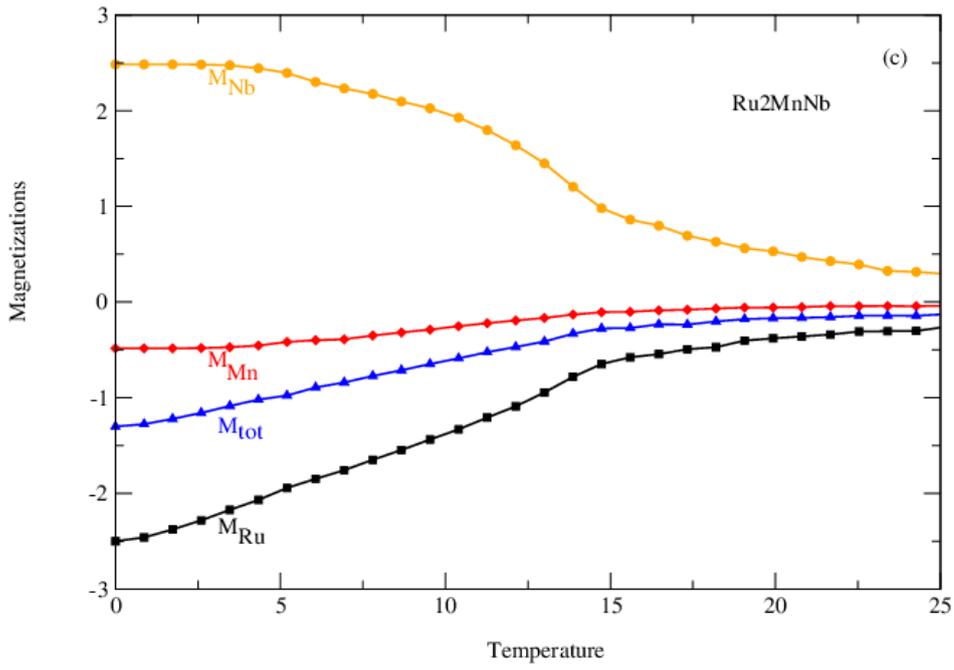

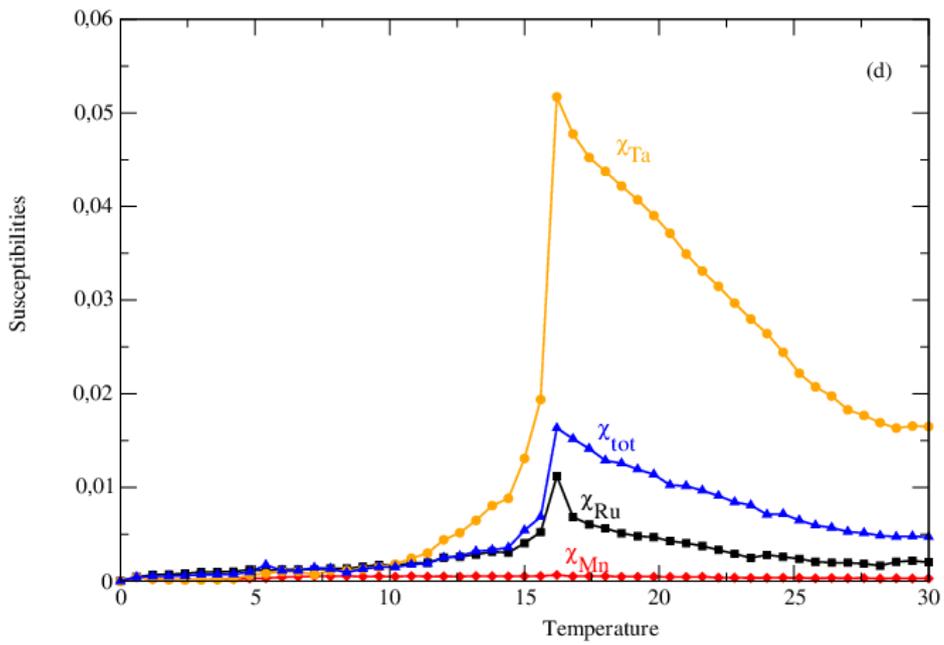

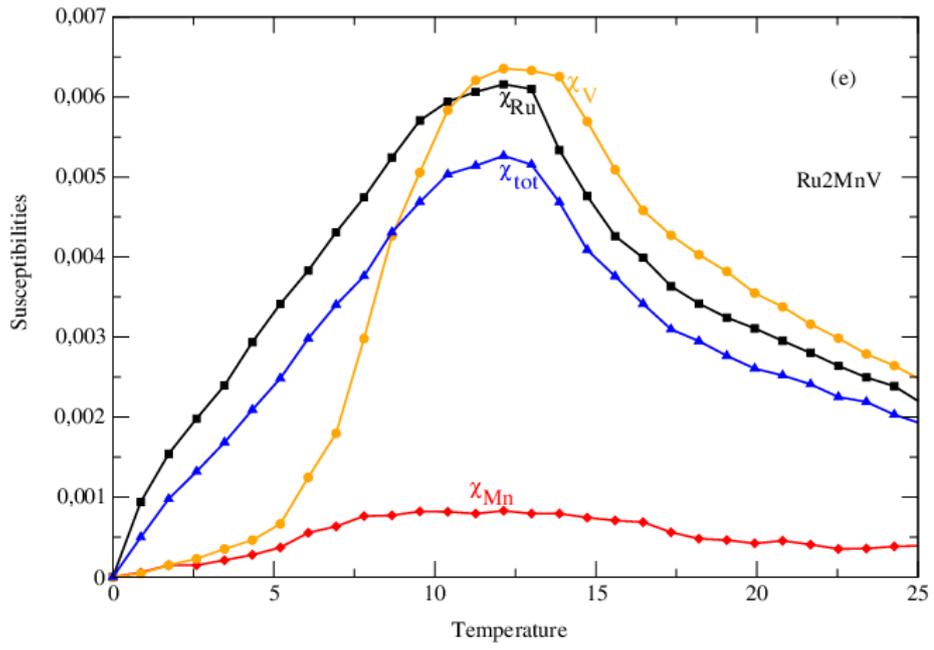

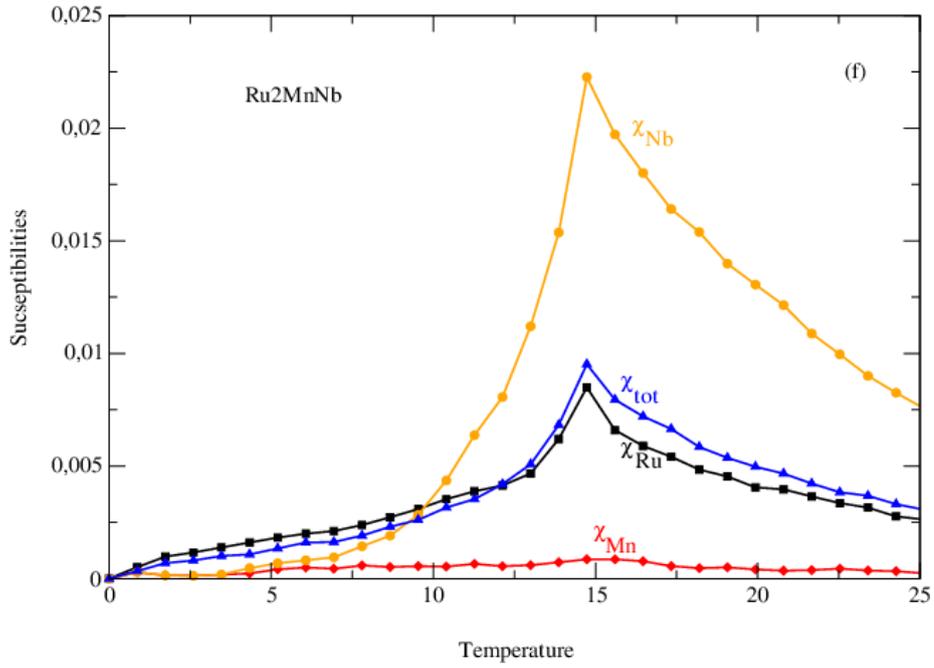

**Fig. 2**: Magnetizations (a, b, c) and susceptibilities (d, e, f) as a function of temperature for fixed parameter values: H=0, D=0, $J_{Ru-Mn}$ = 1 and $J_{Ru-X}$ = -1.

In Fig. 3a, we examine the variation of the thermal magnetization in different nano-Heusler components ($Ru_2MnTa$, $Ru_2MnV$ and $Ru_2MnNb$). This figure is plotted in the absence of the crystal (D=0) and external magnetic (H=0) fields and for the fixed exchange coupling interactions: $J_{Ru-Mn}$ = 1 (ferrimagnetic case) and $J_{Ru-X}$ = -1 (anti-ferrimagnetic case). From Fig. 3a, when increasing the temperature values, the total magnetization appears firstly for the $Ru_2MnV$ compound followed by the $Ru_2MnNb$ compound and finally for the $Ru_2MnTa$ one. This means that $Ru_2MnV$ compound reaches the superparamagnetic phase earlier than $Ru_2MnNb$ compound and ultimately followed by the $Ru_2MnTa$ one. To deduce the blocking temperature ($T_B$) value for each Heusler compounds, we plot in Fig. 3b the thermal susceptibility for the same parameters as in Fig. 3a. The obtained blocking temperature for the $Ru_2MnV$, $Ru_2MnNb$ and $Ru_2MnTa$ compounds are $T_B \approx$ 11, 15 and 18, respectively.

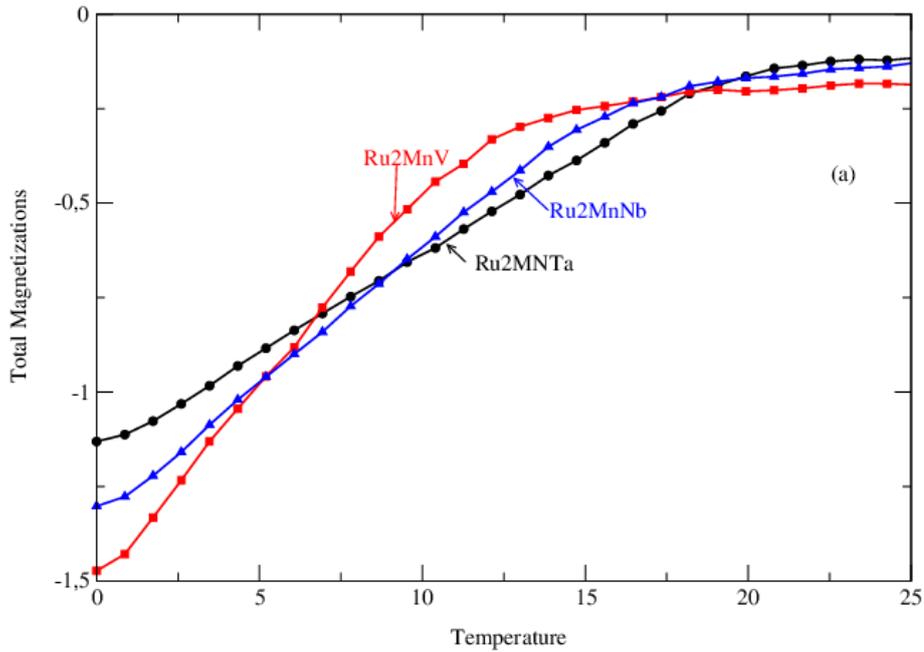

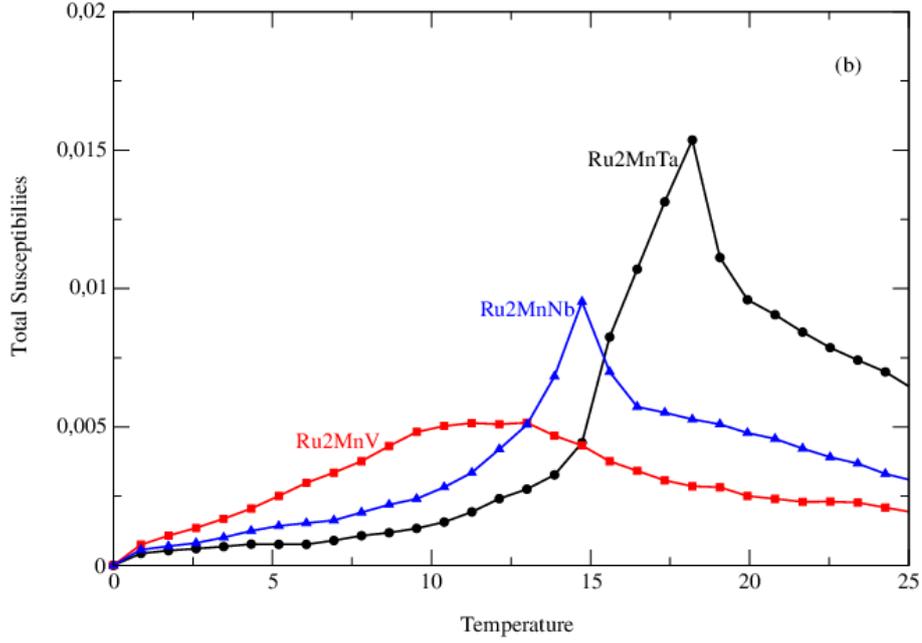

**Fig. 3**: Total magnetizations (a) and total susceptibilities (b) as a function of temperature for fixed parameter values: H=0, D=0, $J_{Ru-Mn}$ = 1 and $J_{Ru-X}$ = -1.

Following the same motivation, we explore in Fig.4a and 4b, the variation of the thermal magnetization in different nano-Heusler components. These figures are plotted in the absence of both fields (H=0 and D=0) and for the fixed ferromagnetic exchange coupling: $J_{Ru-Mn}$ = 1 and $J_{Ru-X}$ = 1. The blocking temperature for the $Ru_2MnV$, $Ru_2MnNb$ and $Ru_2MnTa$ compounds are $T_B \approx$ 12, 17.5 and 20, respectively. The super-paramagnetic phase is observed as the first one for $Ru_2MnV$ compound when (T > $T_B \approx$12) then for $Ru_2MnNb$ compound when (T > $T_B \approx$17.5) and finally for $Ru_2MnTa$ one when (T > $T_B \approx$20). Indeed, the difference between this result and that of Figs. 3a and 3b, is that when increasing the exchange coupling interaction ($J_{Ru-X}$) value, the blocking temperature also tends to increase, which means that the super-paramagnetic phase appears only for large temperature values.

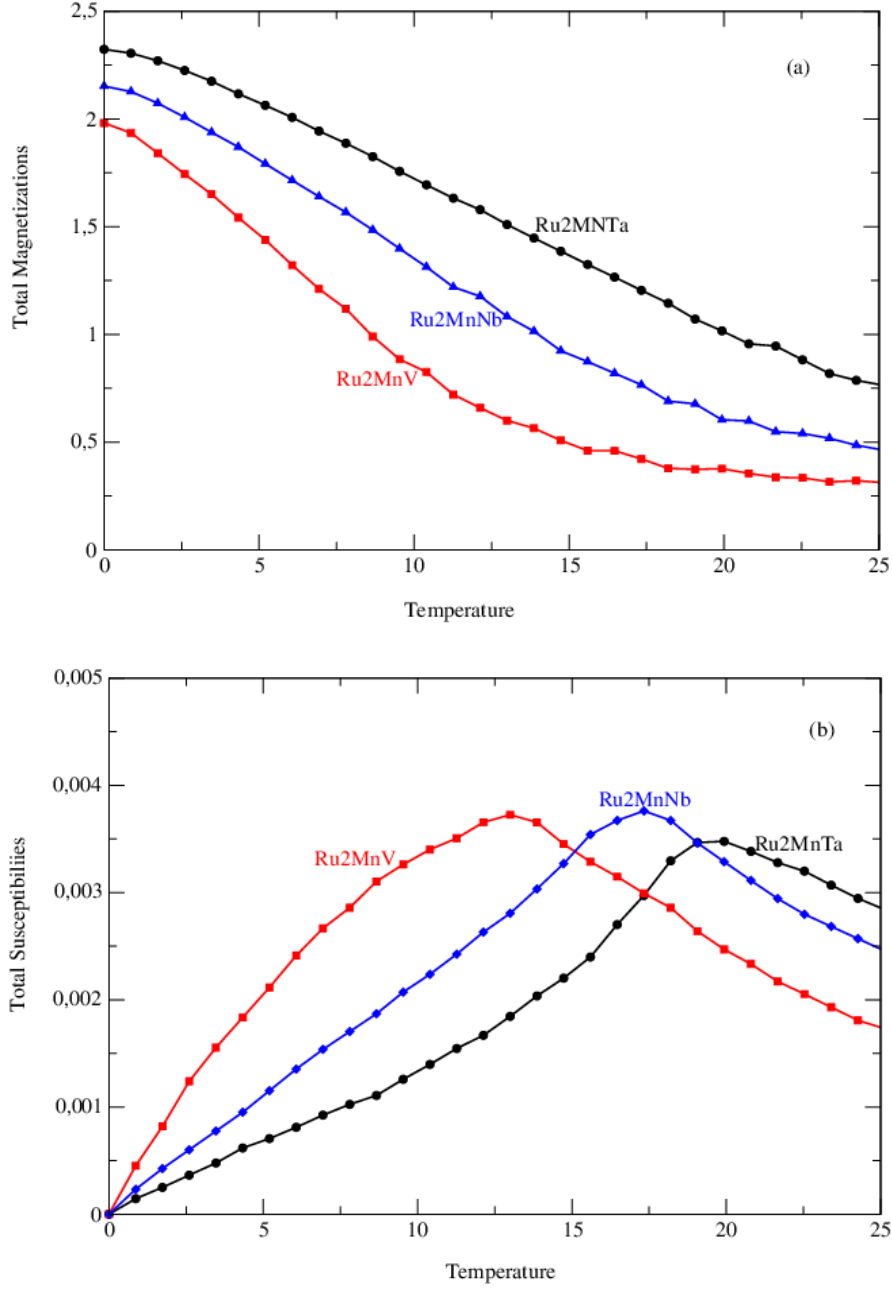

**Fig. 4**: Total magnetizations (a) and total susceptibilities (b) as a function of temperature for fixed parameter values: H=0, D=0, $J_{Ru-Mn} = 1$ and $J_{Ru-X} = 1$.

Besides, we present in the Figs. 5(a-d), the behavior of the total magnetizations and total susceptibilities as a function of the crystal field. Such figures are plotted for the three nan-Heusler compounds ($Ru_2MnV$, $Ru_2MnNb$ and $Ru_2MnTa$) in the absence of the reduced external magnetic field (H=0) and fixed parameter values of temperature T=1 and exchange coupling parameter $J_{Ru-Mn} = 1$ and by changing the exchange coupling interaction between Ru and X atoms: $J_{Ru-X} = -1$ for (a, b), $J_{Ru-X} = 1$ for (c, d).

From Fig.5a, the total magnetizations of the three nano-Heusler compounds undergo a first order transition corresponding to the critical crystal field ($D_T$) values obtained by the susceptibility peaks in Fig. 5b. The obtained critical crystal field for $Ru_2MnV$, $Ru_2MnNb$ and $Ru_2MnTa$ compounds are $D_T \approx$ -7.5, -5.5 and -4, respectively. Then, the total magnetization of these three Heusler compounds is saturated for D >0. In Fig.5c, when increasing the exchange coupling interaction ($J_{Ru-X}$) value, the total magnetizations of the three nano-Heusler compounds undergo a second order transition corresponding to the critical crystal field values given by the susceptibility peaks exposed in Fig. 5d. The obtained critical crystal field for $Ru_2MnNb$, $Ru_2MnTa$ and $Ru_2MnV$ compounds are $D_T \approx$ -6, -4 and -2.5, respectively. Afterward, the total magnetization of these three Heusler compounds is saturated for D >1.

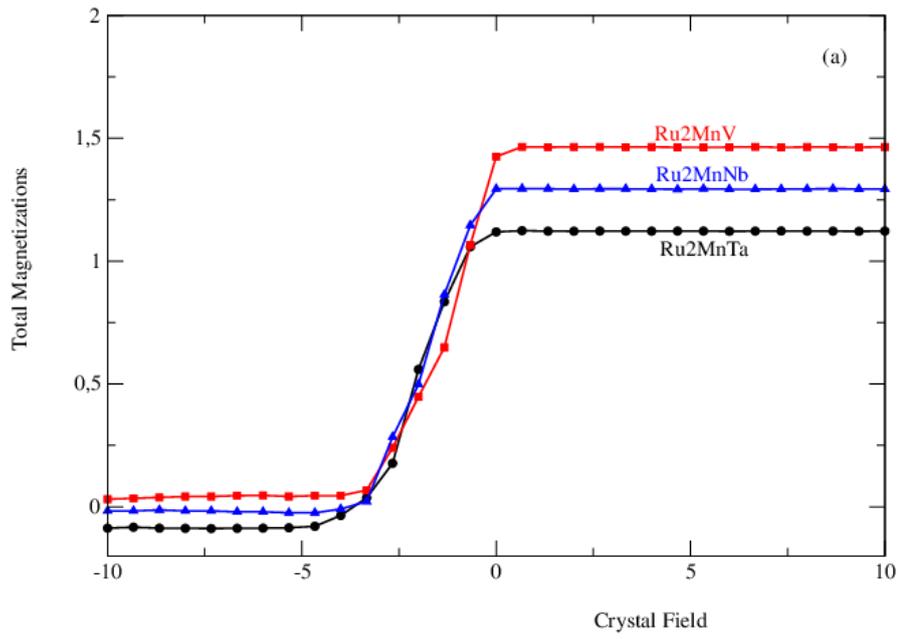
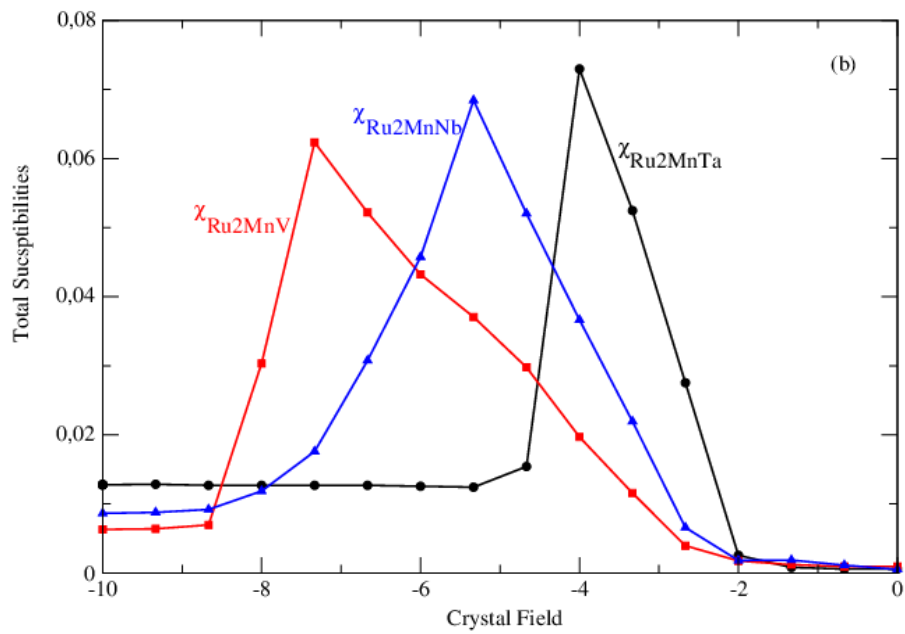

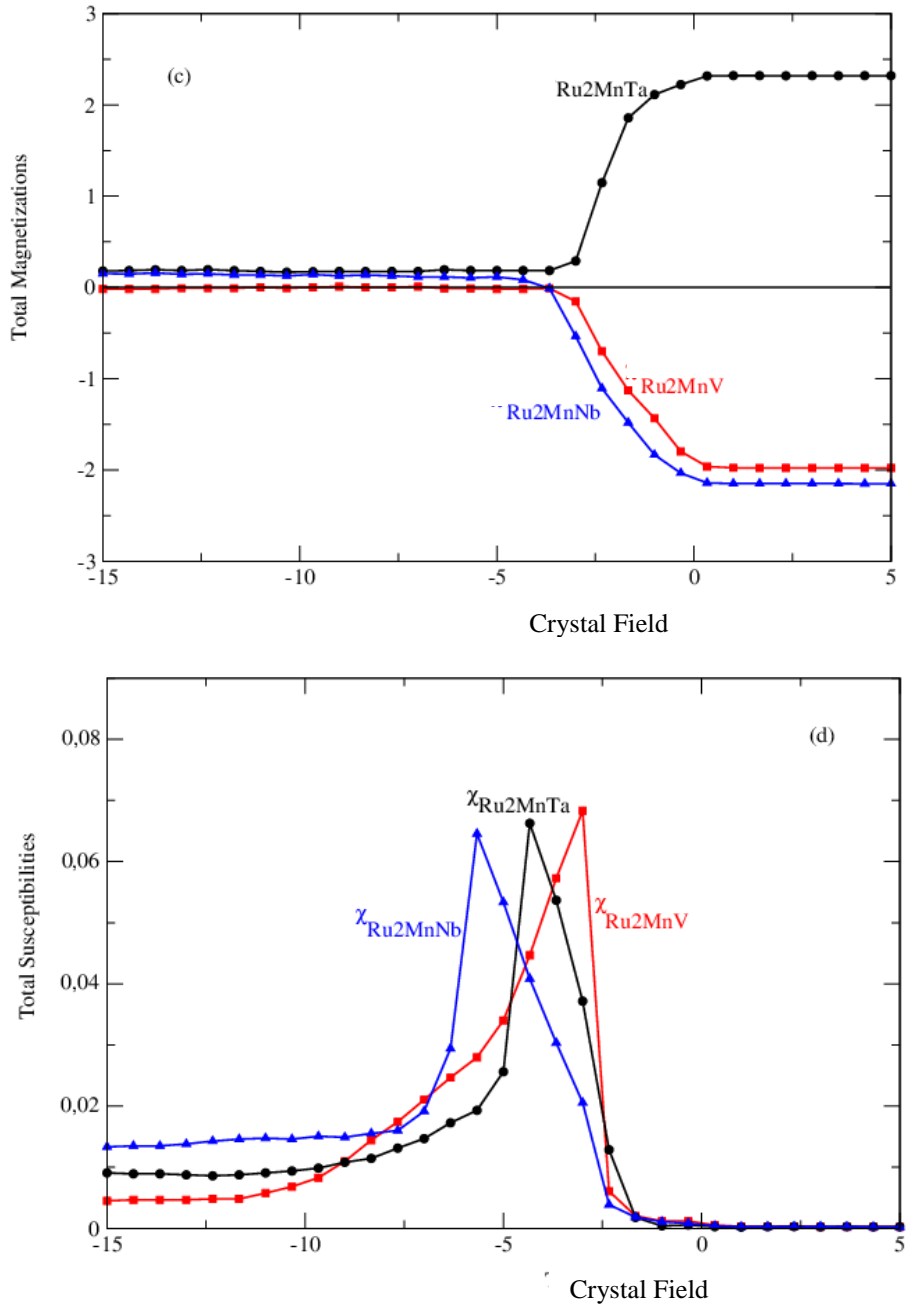

**Fig. 5**: Total magnetizations and total susceptibilities as a function of crystal field for fixed parameter values: H=0, T=1 and $J_{Ru-Mn} = 1$: $J_{Ru-X} = -1$ for (a, b), $J_{Ru-X} = 1$ for (c,d).

Finally, we explore in Figs.6a and 6b the magnetic hysteresis cycle behaviors of the three nano-Heusler compounds. These Figures are plotted in the absence of crystal field (D=0) and for a fixed temperature T=1: Fig. 6a is plotted for $J_{Ru-X}$ = -1 (anti-ferrimagnetic case), while Fig. 6b is illustrated for $J_{Ru-X}$ = 1 (ferrimagnetic case). In Fig.6a, the surface of the loops is important in the $Ru_2MnTa$ compound followed by that of the $Ru_2MnNb$ compound then by that of the $Ru_2MnV$ structure. The obtained coercive field for the $Ru_2MnTa$, $Ru_2MnNb$ and $Ru_2MnV$ compounds are $H_C \approx$ 5, 2.5 and 1, respectively. The decrease in the coercive field and the surface of the loops is due to the difference in the number of atoms and spins from one nano-Heusler compound to another. Furthermore, the behavior of the total magnetization of the nano-Heusler compounds in Fig.3a confirms this result. Moreover, following the same motivation, we present Fig.6b. From this figure, when increasing the exchange coupling parameter $J_{Ru-X}$, the coercive field and the surface of the loops also tends to increase. This behavior confirms the result already found in Fig4a.

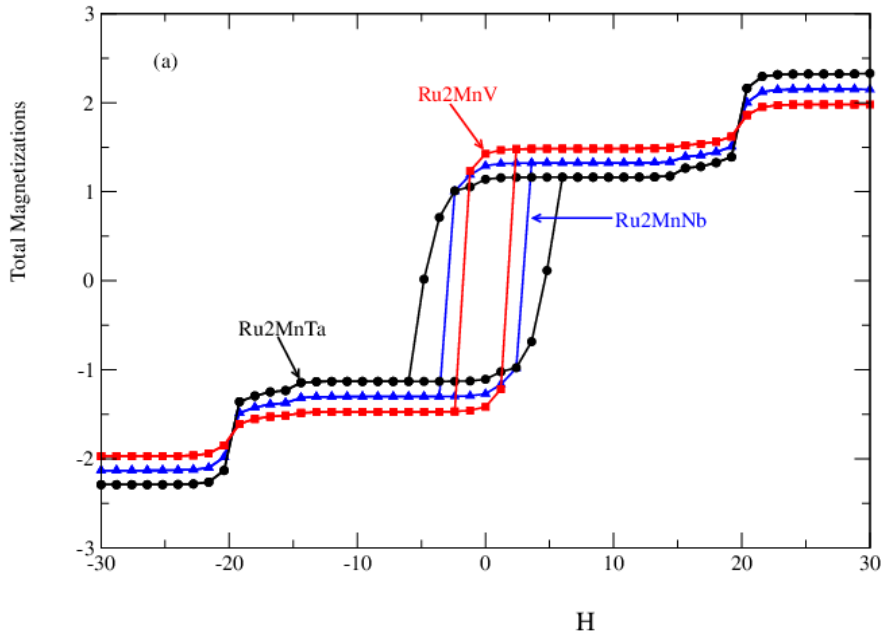

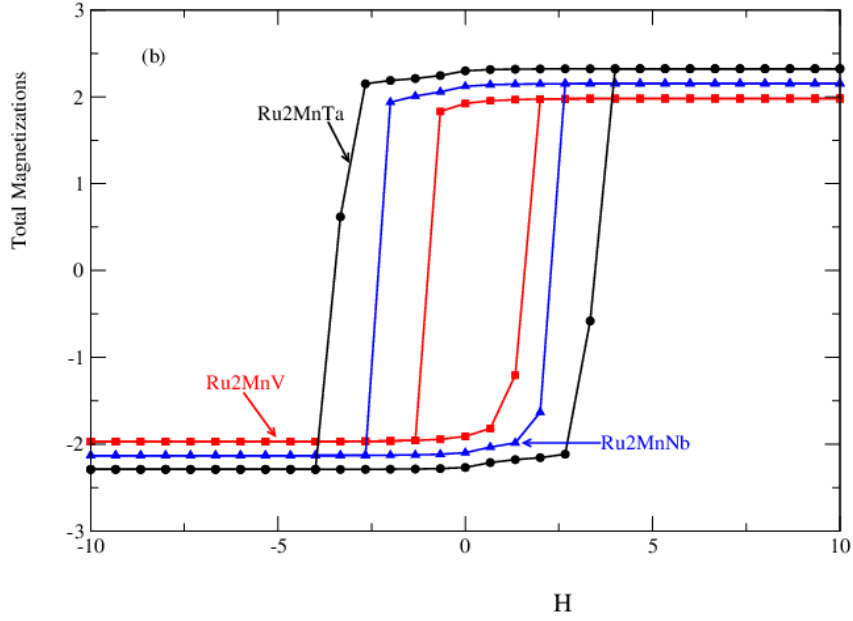

**Fig. 6**: Hysteresis cycles for different Heusler structure for fixed parameter values: D=0, T=1: $J_{Ru-X} = -1$ for (a), $J_{Ru-X} = 1$ for (b).

## 4. Conclusion

In summary, the magnetic properties, including the magnetizations, the susceptibilities, and the blocking temperature $T_B$, of a three nano-Heusler compounds $Ru_2MnX$ (X = Nb, Ta or V) have been studied, using Blume-Capel Ising model using the Monte Carlo simulations. The Ru and Mn atoms are modeled by spin-5/2 and spin-1/2, respectively. While, the X atoms (X = Nb, Ta and V) are represented by the spin-7/2, spin-3/2 and spin-5/2, respectively.

It is found that $Ru_2MnV$ compound reaches the super-paramagnetic phase earlier than $Ru_2MnNb$ compound and ultimately followed by the $Ru_2MnTa$ one. In fact, when increasing the exchange coupling interaction $J_{Ru-X}$ value, the blocking temperature increase which mean that the super-paramagnetic phase appears for larger temperature values. Besides, corresponding to the critical crystal field ($D_T$) values, the total magnetizations of the three nano-Heusler compounds undergo a first order transition or second order transition (when increasing the parameter $J_{Ru-X}$ value). Finally, the surface of the loops and the coercive field are important in the $Ru_2MnTa$ compound followed by the $Ru_2MnNb$ compound and less important in the $Ru_2MnV$ one. Furthermore, when increasing the exchange coupling parameter $J_{Ru-X}$, the coercive field and the surface of the loops also tends to increase.